\documentclass[12pt]{iopart}
\expandafter\let\csname equation*\endcsname\relax
\expandafter\let\csname endequation*\endcsname\relax
\usepackage{mathtools} 
\usepackage[utf8x]{inputenc}
\usepackage{epsfig}
\usepackage{color}
\usepackage[toc,page]{appendix}
\usepackage{float}
\usepackage{graphicx}
\usepackage{anysize}
\usepackage{xfrac}
\usepackage{soul}
\soulregister\cite7
\soulregister\ref7
\soulregister\pageref7

\usepackage{soul}

\marginsize{3cm}{3cm}{2.5cm}{4.5cm}
\begin{document}

\title{Langevin equation in systems with also negative temperatures} 

\author{Marco Baldovin}
\address{Dipartimento di Fisica, Universit\`a di Roma Sapienza, P.le Aldo Moro 2, 00185, Rome, Italy}
\author{Andrea Puglisi~\footnote{Author to whom any correspondence should be addressed.}}
\address{Istituto dei Sistemi Complessi - CNR and Dipartimento di Fisica, Universit\`a di Roma Sapienza, P.le Aldo Moro 2, 00185, Rome, Italy}
\author{Angelo Vulpiani}
\address{Istituto dei Sistemi Complessi - CNR and Dipartimento di Fisica, Universit\`a di Roma Sapienza, P.le Aldo Moro 2, 00185, Rome, Italy }
\address{Centro Linceo Interdisciplinare ``B. Segre'', Accademia dei Lincei, Rome, Italy}
\ead{andrea.puglisi@roma1.infn.it}

\begin{abstract}
We discuss how to derive a Langevin equation (LE) in non standard
systems, i.e.  when the kinetic part of the Hamiltonian is not the
usual quadratic function. This generalization allows to consider also
cases with negative absolute temperature. We first give some
phenomenological arguments suggesting the shape of the viscous drift,
replacing the usual linear viscous damping, and its relation with the
diffusion coefficient modulating the white noise term.  As a second
step, we implement a procedure to reconstruct the drift and the
diffusion term of the LE from the time-series of the momentum of a
heavy particle embedded in a large Hamiltonian system. The results of
our reconstruction are in good agreement with the phenomenological
arguments. Applying the method to systems with negative temperature,
we can observe that also in this case there is a suitable Langevin
equation, obtained with a precise protocol, able to reproduce in
a proper way the statistical features of the slow variables.  In other
words, even in this context, systems with negative temperature do
not show any pathology.
\end{abstract}


\maketitle
\section{Introduction}

It is difficult to overestimate the relevance of the Langevin equation
(LE) which is one of the few pillars of the non equilibrium
statistical mechanics \cite{vanKampen2007,livipoliti}.  About a
century ago, in his seminal paper on Brownian motion, Langevin
introduced his celebrated stochastic differential equation with the
aim to describe the long time statistical features of a colloidal
particle~\cite{langevin,lemons}.  The derivation was phenomenological,
namely based upon macroscopic arguments (the Stokes law), and
statistical assumptions (thermal equilibrium of the colloidal particle
with the liquid).  The LE had an important role in mathematics: the
work of Langevin had been the starting point for the building of a
general stochastic process theory~\cite{vanKampen2007,gardiner}.

A natural question is the possibility to derive the LE in a non
phenomenological way, i.e.  starting from the dynamics of large
systems \cite{zwanzig}.  Unfortunately there are just few cases where
it is possible to use such a desirable
approach~\cite{castiglione2008}.

One situation is the motion of a heavy particle in a diluted gas: in
such a case, using an approach going back to Smoluchovski, with a
statistical analysis of the collisions among the heavy particle and
the light gas particles, it is possible to determine the viscous
friction~\cite{von1906kinetischen,cecconi2007transport}. A complete
derivation, including the shape of the noise term, has been also
obtained as a perturbative expansion of the Lorentz-Boltzmann equation
by van Kampen~\cite{vK61}, repeated in a similar form for granular
gases~\cite{sarracino}.  There is also another large class of systems
where it is possible to obtain a LE in an analytical way: harmonic
chains with a heavy particle of mass $M$ and $N$ light particles of
mass $m \ll M$~\cite{rubin60,turner60,zwanzig73}.  In such a case the
linearity of the dynamics allows for an explicit solution and then the
possibility to find the LE for the heavy particles in the limit $m/M
\ll 1$ and $N \gg 1$. As far as we know there are no other clearly
distinct cases where it is possible to derive a LE of an heavy
particle interacting with many light particles with an analytical
approach, i.e. starting from the deterministic dynamics of the whole
system. 

In the last years the LE played a prominent role in the statistical
mechanics of small systems, i.e. those containing a limited number of
particles, in particular for the stochastic thermodynamics
approach~\cite{Seifert,livipoliti}.  Therefore it seems highly
desirable to have the possibility to write down a LE for a heavy
particle also in systems different from the well known cases above
discussed. Approximate derivations for the case of non-linear forces~\cite{maes17} and
for cases near non-equilibrium stationary processes~\cite{maes15} have also been recently
considered. A particularly interesting test-case is that of systems
where the Hamiltonian has a non-standard kinetic term,
i.e. non-quadratic in the momentum, leading to non-trivial properties
such as negative absolute
temperature~\cite{cpv,tempreview,termometri}, a possibility
recently verified also in experiments~\cite{Braun2013}.

There are different physical mechanisms which allow for the validity
of a LE: this becomes clear when comparing the dynamics of a colloidal
particle in a dilute fluid with that of a heavy mass in an harmonic
solid. In most cases a separation of time scales is a necessary
condition which is guaranteed by the condition $M \gg m$. Often this
is a strong indication of the validity of a LE, with possible
exceptions discussed at the end of Section 3. In most of the present paper we
simply assume a separation of time scales between the slow and the
fast variables (with a few cases where we verify it), therefore our approach is merely phenomenological
and numerical~\cite{peinke,peinke2,kantz,engquist,givon}. Some authors
have discussed the origin of such an assumption, within a dynamical
systems approach~\cite{kifer,mackay}.

The aim of the present paper is twofold: a) the introduction of a
practical procedure, which can be used also with data from
experimental results, to build a Langevin equation from a long time
series; up to our knowledge similar procedures have been applied,
  previously, to model systems of different kinds (e.g. turbulence) but
 never to Hamiltonian systems; b) to show that
systems characterized by negative temperatures does not display any
pathology, i.e. that in such a kind of systems, for the dynamics of slow
variables one can adopt a consistent efficient description in terms of
a Langevin equation whose parameters can be computed with a well
defined protocol.

The structure of the paper is the following. In Section 2 we review
some general aspects of Langevin equation, and present
phenomenological arguments to determine the shape of the viscous term
in the presence of additive noise.  Section 3 is devoted to the
numerical procedure of building Langevin equation from a long
time series.  The comparison of the actual results obtained from a
numerical simulation and the predictions from Langevin equation
are in very good agreement.  Such a consistency holds in the presence
of a time scale separation between the slow variable and the ``bath''
(i.e. the rest of the system) while the non standard shape of the
(generalized) kinetic energy in the Hamiltonian does not play any crucial
role.  In particular we have that in the cases with negative
temperature there is no pathology, and Langevin equation is
able to reproduce the expected statistical features.  In Section 4 we
draw some concluding remarks.  In the Appendix we present important details of
the numerical procedure.

\section{General considerations}
\label{general}

Let us start with discussing some general features of the LE and then
present phenomenological arguments to guess the shape of the friction
term of a probe (e.g. ``heavy'') particle in a system with a
non-quadratic ``kinetic energy''.

\subsection{Model and Langevin description}

Consider a Hamiltonian system of the form
\begin{equation} \label{ham_gen}
H(P,\{p\},Q,\{q\})=K(P)+\sum_n \tilde{K}(p_n)+U(Q)+\sum_n V_I(Q,\{q_n\})+\sum_{n,n'}\tilde{V}(q_n,q_n')
\end{equation}
where $(P,Q)$ denote the canonical variables of a ``heavy'' particle
and $(\{p_n\},\{q_n\})$ indicate the ``light'' particles. We have
denoted with $\tilde{V}$ and $V_I$ the potential for the interaction
between light particles and that for the interaction heavy-light particles,
respectively, while $U$ is the external potential confining the heavy
particle; $K$, $\tilde{K}$ are the kinetic energies of the heavy
and light particles respectively.

The evolution equation for $(P,Q)$ are
\begin{subequations}
  \begin{align}
    \dot{Q}&=\partial_P H=\partial_P K(P)\\
    \dot{P}&=-\partial_Q H=-\partial_Q U(Q)-\sum_n \partial_Q V_I(Q,\{q_n\}). \label{dotP}
  \end{align}
\end{subequations}
Under the hypothesis of time-scale separation, as in the cases
of massive particles discussed in the Introduction, one expects that the term $-\sum_n
\partial_Q V_I$ can be described by a ``viscous term'' - only function
of the variables $P,Q$ - and a noisy term. This amounts to look for a generalization of the ``Klein-Kramers'' equation for
a generic form of $K(P)$, i.e. including cases where the kinetic term is different from $P^2$ and therefore one
may have also ranges with inverse temperature $\beta<0$. Our candidate is an equation of the kind:
\begin{subequations} \label{KK}
\begin{align}
  \dot{Q} &= \partial_P K(P)\\
  \dot{P} &= -\partial_Q U(Q) + B(P,Q,t), \label{dotp}
\end{align}
\end{subequations}
where $B(P,Q,t)$ is the effective force due to the interaction with
the rest of the system, such as a thermal bath in standard
cases. One may wonder how the dependence upon $P$ 
  in the last term of Eq.~\eqref{dotp} emerges from manipulating the
  last term of Eq.~\eqref{dotP} which only depends upon $Q$ and
  $\{q_n\}$. Even without giving mathematical details (which are
  different for each particular case), it is easy to understand that
  the procedure from Eq.~\eqref{dotP} to Eq.~\eqref{dotp} includes
  conditional averages over the fast degrees of freedom, keeping fixed
  $P$: this implies that the statistical properties of some
  coarse-grained force representing $\sum_n \partial_Q V_I(Q,q_n)$
  must necessarily depend upon the value of $P$.

In this section we assume that $B(P,Q,t)$ takes the simplified form:
\begin{equation} \label{KK2}
  B(P,Q,t) = \Gamma(P)+\sqrt{2D_P}\xi(t), 
\end{equation}
where $\xi(t)$ is a Gaussian white noise, with $\langle \xi(t) \rangle=0$ and
$\langle \xi(t)\xi(t') \rangle = \delta(t-t')$ and e $D_P > 0$. The relaxation of such a hypothesis is described at the end of Section 3.

\subsection{Overdamped case}

Here we show a first argument to determine the function
$\Gamma(P)$. If one requires that the inertial term $\dot{P}$ can be neglected, the only way to have a closed equation for $\dot{Q}$ is to impose
\begin{equation}
\Gamma(P)= c \partial_P K(P),
\end{equation}
with  $c$ some constant to be found. With such a choice, in fact, by means of setting to $0$ the left hand side of Eq.~\eqref{dotp} one gets
\begin{equation}
\dot{Q} = \partial_P K(P)= \frac{\Gamma(P)}{c} = \frac{1}{c}\partial_Q U(Q)-\frac{\sqrt{2D_P}}{c}\xi(t),
\end{equation}
which has the steady probability density $f_Q(Q) \sim
\exp[cU(Q)/D_P]$. Such a density must be consistent with equilibrium, which implies $c=-\beta
D_P$. In summary one has
\begin{subequations} \label{over}
\begin{align}
  f_Q(Q) &\sim \exp[-\beta U(Q)], \\
  \Gamma(P)&=-D_P\beta\partial_P K(P).
\end{align}
\end{subequations}

\subsection{Case with inertia}

The Fokker-Planck equation associated to Eqs.~\eqref{KK}-\eqref{KK2}, in the steady state reads
\begin{align} \label{fok}
  \partial_Q J_Q(Q,P)+\partial_P J_P(Q,P)&=0\\
  J_Q(Q,P)&=f(Q,P)\partial_P K(P)\\
  J_P(Q,P)&=-f(Q,P)\partial_Q U(Q,P)+\Gamma(P) f(Q,P) - D_P\partial_P f(Q,P),
\end{align}
where $f(Q,P)$ is the steady probability density.

A solution for $f(Q,P)$ can be found by asking that detailed balance
is satisfied, as it must occur at thermodynamic equilibrium~\cite{gardiner}. This
condition is equivalent to ask that the part of $J_P$ associated to
the thermal bath (the so-called ``irreversible current'') vanishes,
i.e.:
\begin{equation}
\Gamma(P) f(Q,P)-D_P \partial_P f(Q,P)=0,
\end{equation}
which can be solved by factorization, i.e. $f(Q,P) =f_Q(Q) f_P(P)$, leading to $f_P(P) \sim \exp[-\beta K(P)]$ and to Eqs.~\eqref{over}.


\subsection{Discussion}


Of course in the most common case, i.e. when $K(P)=P^2/(2 M)$, one
recovers $\Gamma(P)=-D_P\beta P/M$, that is the usual viscous term
$-\gamma V$ with viscosity satisfying the Einstein relation
$\gamma=\beta D_P$ and therefore it can only be $\beta>0$.
Interestingly, one always has $\Gamma(P)=-D_P\beta \dot{Q}$,
i.e. somehow the ``velocity'' $\dot{Q}$ sees no consequences of the
different shape of the kinetic term $K(P)$.  Moreover, in all cases
one obviously has $f_P(P) \sim \exp [-\beta K(P)]$. It is clear that
in cases where $\beta<0$ boundary conditions on $P$ must be consistent
with the normalization of $f_P(P)$.

In the model for negative temperatures discussed in Section 3~\cite{cpv}, one has $K(P)=1-\cos(P)$ and
therefore $\Gamma(P)=-D_P\beta \sin(P)$, which let $\beta$ have any
possible sign. It is interesting to notice that the ``drift'' term
$\Gamma(P)$ acts consistently with the simple idea deduced from the
form of $f_P(P)$: the drift term should counteract the spreading
action of the noise term in order to concentrate the distribution in
its maximum. Indeed, when $\beta>0$ such a distribution is peaked
around $P=0$ and the drift pushes $P$ far from $P=\pm \pi$ and towards
$P=0$. On the contrary, when $\beta<0$ the distribution is peaked near
$P=\pm \pi$ and in fact the drift pushes $P$ far from $P=0$.


\section{Empirical procedure to determine the parameters of a Langevin equation}
\label{sec:ham}

Almost all important problems in science are characterized by the
presence of a variety of degrees of freedom with very different time
scales. Among the many examples we can mention protein folding and
climate: for proteins, the time scale of the vibration of covalent bonds is
$O(10^{-15})$ s, while the folding time may be of the order of
seconds; in the case of climate, the characteristic times of the
involved processes vary from days (for the atmospheric phenomena) to $O(10^3)$ yr
for the deep ocean flows and ice shields.

The necessity of treating
the ``slow dynamics'' in terms of effective equations is both practical
(even modern supercomputers are not able to simulate all the relevant
scales involved in certain difficult problems) and conceptual:
effective equations are able to catch some general features and to
reveal dominant ingredients which can remain hidden in the detailed
description. The study of such multiscale problems has a long history in
science, and some very general mathematical methods have been
developed~\cite{castiglione2008, engquist, givon}, whose usage, however, is often not easy at all.
In the present paper we adopt, instead, a rather practical numerical procedure to
build the Langevin equation: such an approach is quite natural and it has been
already used in the study of turbulence~\cite{kantz,peinke_comment}.

In what follows we will consider three Hamiltonian systems, with
different kinetic terms, and exhibit a constructive procedure to infer
the Langevin parameters of a slow degree of freedom \textit{a
  posteriori}, i.e. analyzing the data produced by the deterministic
molecular dynamics simulations; the outcomes will be then compared to
our predictions of Section~\ref{general}. We must note that such a comparison is only 
possible if the reconstructed noise amplitude is non-multiplicative, as assumed
in the previous Section: if this is the case, we expect to verify
Eq.~\eqref{over}. We stress, however, that the procedure described in this
Section is also valid if the noise is multiplicative.

%

\begin{enumerate}
\item The first system we consider is a chain of $2N+1$ coupled harmonic oscillators with Hamiltonian
\begin{equation} \label{ham_osc}
 H=\frac{P^2}{2M} + \sum_{i=\pm 1, ..., \pm N} \frac{p_i^2}{2m} +  \frac{k}{2}\sum_{i=-N}^{N+1} (q_i-q_{i-1})^2, \quad \quad q_{-N-1}\equiv q_{N+1}\equiv 0
\end{equation}
 in which the heavy particle, that will be referred to as the
 \textit{intruder}, occupies the central position ($Q\equiv q_0$). $k$
 here represents the elastic constant, while $m$ and $M$ are the mass
 of the light particles and that of the intruder, respectively. We
 adopt fixed boundary conditions for the first and the last particles
 for computational reasons: they prevent an unbounded drift of
 positions caused by the conservation of total momentum.  \\ This system can be solved analytically; in a slightly
 modified version it has been extensively studied since 1960, when
 Rubin and Turner, in their seminal works \cite{rubin60,turner60},
 showed that the behavior of the heavy particle could be approximated
 by a Brownian motion, under the assumption of canonically distributed
 initial conditions. In particular they were able to prove that the
 autocorrelation function of the heavy particle's velocity $C(t)$
 could be approximated by
 \begin{equation}
  C(t) \sim \exp\left(-\frac{2\sqrt{k m}}{M-m} t \right) + O(m/M)
 \end{equation}
 when $M/m \gg 1$. Further analyses \cite{takeno-hori62} on the frequency
 spectrum of the normal modes pointed out that
 the previous approximation was valid only if the ratio $M/\sqrt{k m}$ continued 
 to be finite when the heavy mass limit was taken. Several generalizations of this simple model
 have been explored: the linear chain with nearest-neighbours interactions
 has been shown to be just a particular case of a wider class of harmonic systems
 with similar properties \cite{mazur-braun64} \cite{ford-kac-mazur65}, that can be 
 used as ``thermal baths'' for the intruder even if the heavy particle is subjected
 to non-linear forces \cite{zwanzig73}.
 \\
Since the properties of this harmonic chain are completely known, checking our method
on this model is a quite natural choice. Further details on the numerical protocol and its
application in this specific case are given in the Appendix.

\item The second model is a slight generalization of the harmonic chain discussed before:
all kinetic terms (including that of the heavy particle) are replaced by $mc^2 f\left(\frac{p}{cm}\right)$,
where $p$ and $m$ represent the momentum and the mass of the considered particle,
$c$ is a characteristic constant with the dimensions of a velocity
and $f(x)$ is an even function of $x$ (the previous situation is recovered for $f(x)=x^2/2$).
This choice for the kinetic energy is due to extensivity requirements: if we ask that
$n$ particles with equal masses, positions and velocities should take the same total kinetic
energy and momentum of a single particle with mass $n m$, such form turns out to be the
only available option. Note that this assumption implies that 
velocity $\dot{q}$ depends on the ratio $p/m$ only (not on $p$ and $m$ separately).
In the following we will consider $f(x)=x^4/4$. Therefore, the Hamiltonian of the system will read:
\begin{equation} \label{ham_quart}
 H=\frac{P^4}{4M^3} + \sum_{i=\pm 1, ..., \pm N} \frac{p_i^4}{4m^3} +  \frac{k}{2}\sum_{i=-N}^{N+1} (q_i-q_{i-1})^2,
 \quad \quad q_{-N-1}\equiv q_{N+1}\equiv 0,
\end{equation}
 where adimensional units have been used and $c$ has been set equal to 1.
We stress that there is no particular reason to choose quadratic
potentials, and the nearest-neighbours interaction is also arbitrary:
only the quality of the agreement with theory constitutes a criterion,
\textit{a posteriori}, to evaluate the limits of our procedure. The only
(important) hint, {\it a priori}, is given by the fact that time-scale
separation should occur when the limit $M/m\gg1$ is taken (with the caveats
discussed in Subsection~\ref{caveats}).

 \item Finally, we would like to test Eq.~\eqref{over} when the system
 admits absolute negative temperatures, i.e. when the derivative of the
 microcanonical entropy with respect to the energy, $\beta \equiv \partial S/\partial E$,
 is allowed to be less than zero. Let us just recall that, at least in the thermodynamic
 limit, negative temperatures can only be achieved when the canonical variables are bounded;
 therefore they cannot be observed on systems with the ``usual'' quadratic kinetic
 energy, and interaction terms also need to be finite~\cite{cpv,iubini2013discrete,Braun2013,tempreview,termometri}. 
 A simple system which satisfies the previous requirements is the following Hamiltonian
 chain, introduced in \cite{cpv}: 
 \begin{equation} \label{ham_neg1}
 H_{chain}=\sum_{i=1}^{N} m[1-\cos(p_i/m)] + k \sum_{i=1}^{N+1} [1-\cos(q_i-q_{i-1})], \quad \quad q_{0}\equiv q_{N+1}\equiv 0.
\end{equation}
 Here $\{p_i,q_i\}$ are the canonical variables of $N$ coupled rotators
 with bounded kinetic energy.
 Note by the way that in Eq.~\eqref{ham_neg1} kinetic terms have
 been written in the form dictated by the extensivity condition seen
 before; also in this case, we have chosen adimensional units in which $c=1$.
 \\
 In order to study the behavior of an additional heavy particle,
 we can couple it to some elements of the chain through bounded
 potential terms similar to the previous ones.  A possible choice is
 given by
  \begin{equation} \label{ham_neg2}
 H=H_{chain} + M[1-\cos(P/M)]+ k \sum_{i=1}^{N/n} [1-\cos(Q-q_{i\cdot n})]
\end{equation}
where $n$ is a positive integer number. In this way the heavy particle is only
linked to rotators labelled as $n, 2n, 3n, ...$. 
One could ask why to choose this kind of coupling, instead of simply replacing 
the central particle with an heavier intruder, as in the previous cases; the reason
is twofold. First, since interaction terms are now bounded, heat exchange between
the various parts of the chain is much slower: a more ``connected'' geometry 
surely enhances the thermalization process. Secondly, it is reasonable that the 
composition of several interactions with different particles of the system 
will result in an uncorrelated noise for the heavy particle, which is a needed
condition for Eq.~\eqref{KK2} to be valid.
Even if this coupling is a quite arbitrary choice, which does not seem to correspond
to any simple experimental setting, we stress that it shares its central features
with much more realistic scenarios: in particular, it allows the heavy intruder to simultaneously
interact with many (weakly correlated) particles of a fast-equilibrating bath (the chain),
providing a minimal schematization for a colloidal suspension.
\end{enumerate}
In the light of the above discussion, as working hypothesis let us assume that in all the previous cases,
in the limit $M/m \gg 1$ and $N \gg 1$, the variable $P$ can be described by a 
Langevin equation with the shape
\begin{equation} \label{langevin}
{d P \over d t}=F(P) + \sqrt{2D(P)} \eta
\end{equation}
where $\eta$  is a white noise with $\langle \eta(t) \eta(t') \rangle =\delta (t- t')$.
In the following we shall see that our assumption holds.
The function $F(P)$, as well as $D(P)$, can be obtained  from a long time series;
it is enough to follow the standard definition in textbooks:
being 
$$
\Delta P(\Delta t)= P(t+\Delta t) - P(t)
$$
we have
\begin{subequations}\label{coeff}
\begin{align}
F(P)&=\lim_{\Delta t \to 0} {1 \over \Delta t} \langle \Delta P(\Delta t) | P(t)=P \rangle\\
D(P)&=\lim_{\Delta t \to 0} {1 \over 2\Delta t} \langle \Delta P(\Delta t)^2 | P(t)=P \rangle \,.
\end{align}
\end{subequations}
\\
These averages can be easily computed from molecular dynamics simulations. We use a
Velocity Verlet algorithm, choosing integration steps $\delta t$ small enough to avoid,
in the various cases, relative energy fluctuations greater than $10^{-4}$: in this
way the momenta of light particles take $\sim 50\, \delta t$ to decorrelate, for all considered
schemes. Further details are in the Appendix.
\\
Some remarks are in order:
\begin{itemize}
\item  The guess that $P$ is described by a Markovian process is quite natural,
and we know that, at least in the harmonic case, it is verified: in the  spirit 
of the works of Rubin and Turner \cite{rubin60, turner60}, also for the systems here considered,
 it is reasonable to assume that the large mass of the intruder 
 with respect to the light particles allows for a separation of time scales, so that  
the momentum  of the impurity is expected to follow a Langevin equation.
\item Even if the guess that $P$ is described by a Markovian process
discussed above can sound quite obvious, actually it is not trivial at all.
The difficulty of the general problem has
been stressed by Onsager and Machlup in their seminal work on
fluctuations and irreversible processes \cite{onsager}, with the caveat: {\it how do
  you know you have taken enough variables, for it to be Markovian?}
In a similar way, Ma noted that \cite{ma85}: {\it the hidden worry of
  thermodynamics is: we do not know how many coordinates or forces are
  necessary to completely specify an equilibrium state.}
\item Surely the approximation of the dynamics of the intruder
as a LE cannot be valid at very small time difference, therefore the limit $\Delta t \to 0$
must be interpreted in a physical way, and it is necessary to fix
a protocol for the computation of $F(P)$ and $D(P)$ from the
 time series $\{ P(t) \}$ with $0<t<{\cal T}$ being ${\cal T}$ large enough:
see the Appendix for a discussion about this point.
\end{itemize}
\begin{figure}
 \centering
 \includegraphics[width=0.49\linewidth]{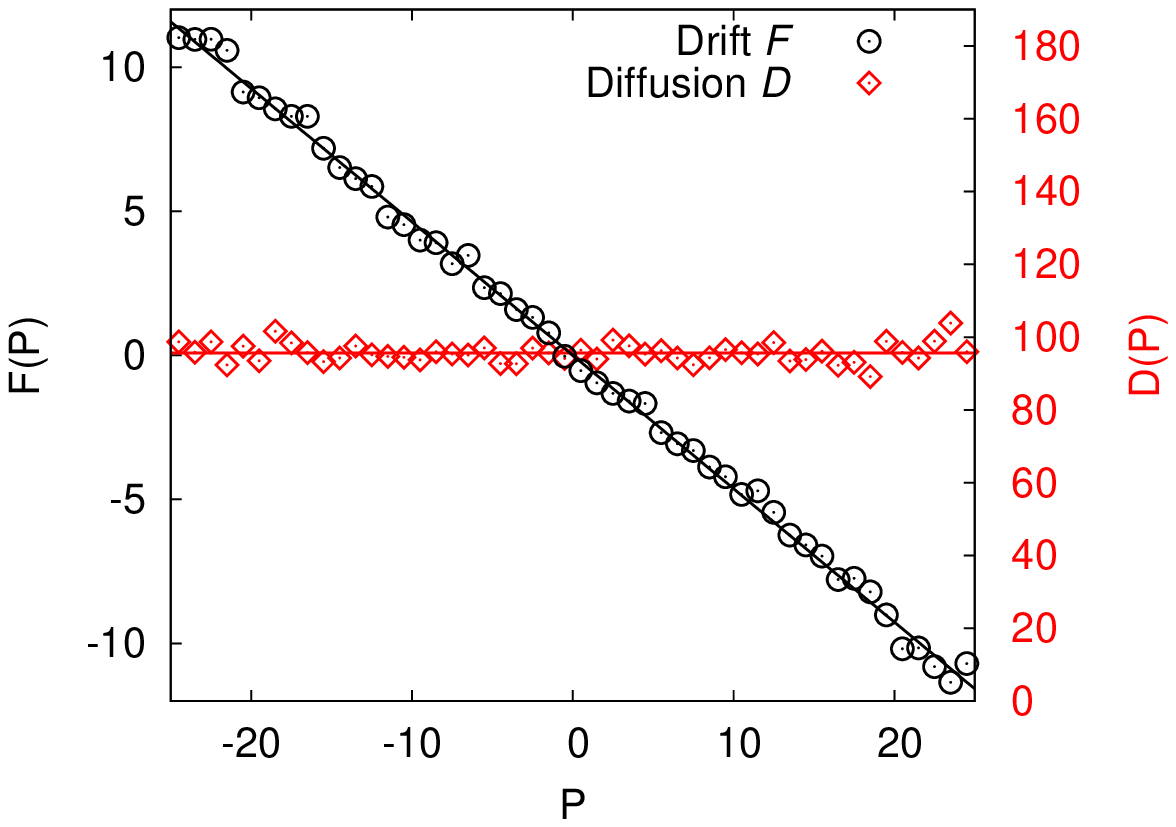}
 \includegraphics[width=0.49\linewidth]{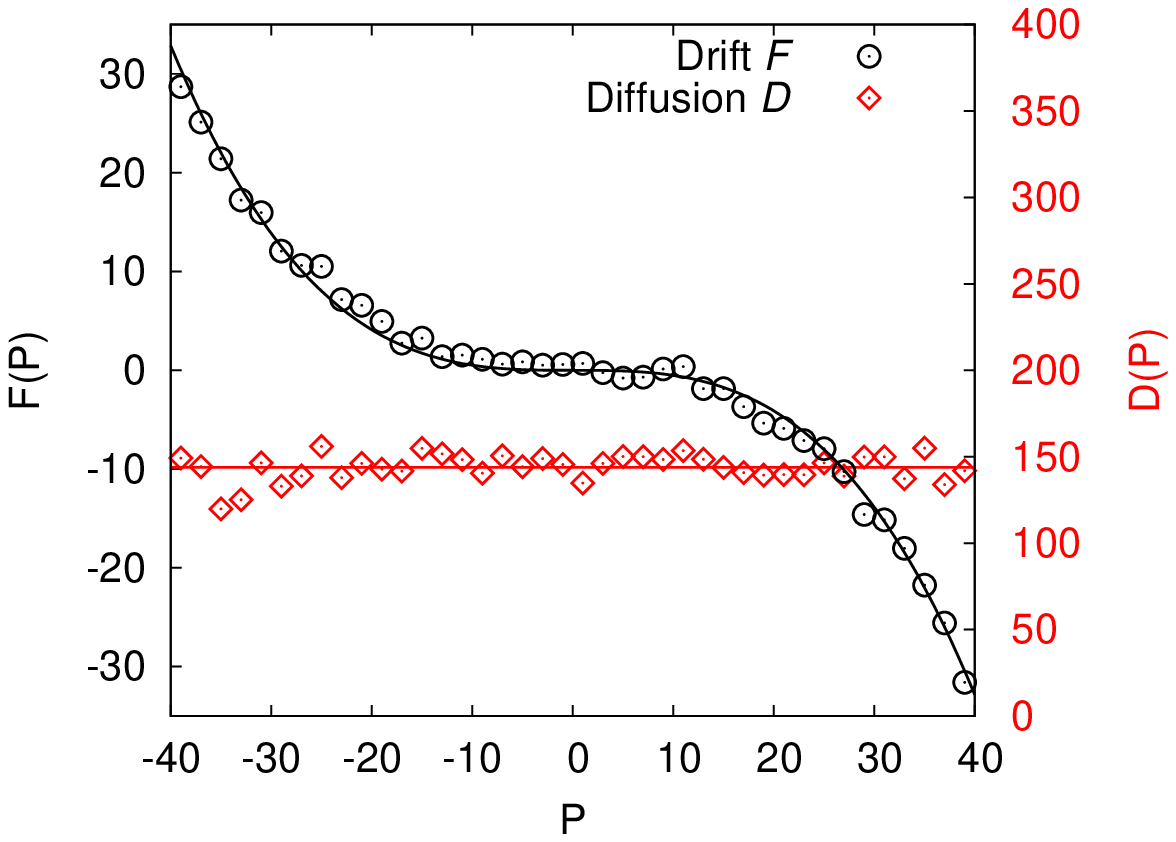}
 \caption{Drift and diffusion coefficients for the Langevin Equation describing $P$,
 inferred from simulations.
 Left: model \eqref{ham_osc}  with $M=200$, $k=2500$, $2N=2000$, $\beta\simeq1.0$;
 integration step $\delta t=10^{-3}$.
 Right: model \eqref{ham_quart} with $M=50$, $k=2500$, $2N=2000$, $\beta\simeq0.45$;
 integration step $\delta t=4\cdot 10^{-4}$.
 Drifts are fitted with functions in the form \eqref{over}, diffusivities with constant values. }
 \label{fig:fit_osc}
 \end{figure}
  \begin{figure}
 \centering
 \includegraphics[width=0.49\linewidth]{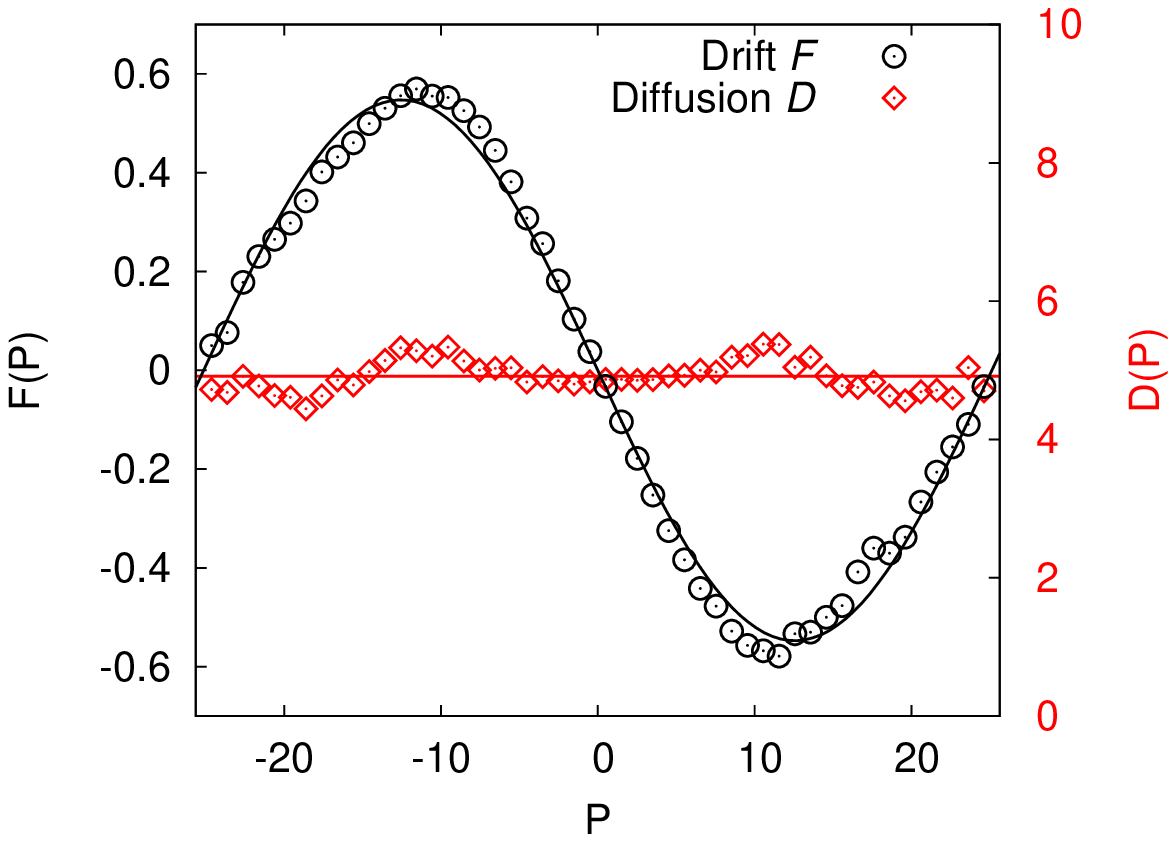}
 \includegraphics[width=0.49\linewidth]{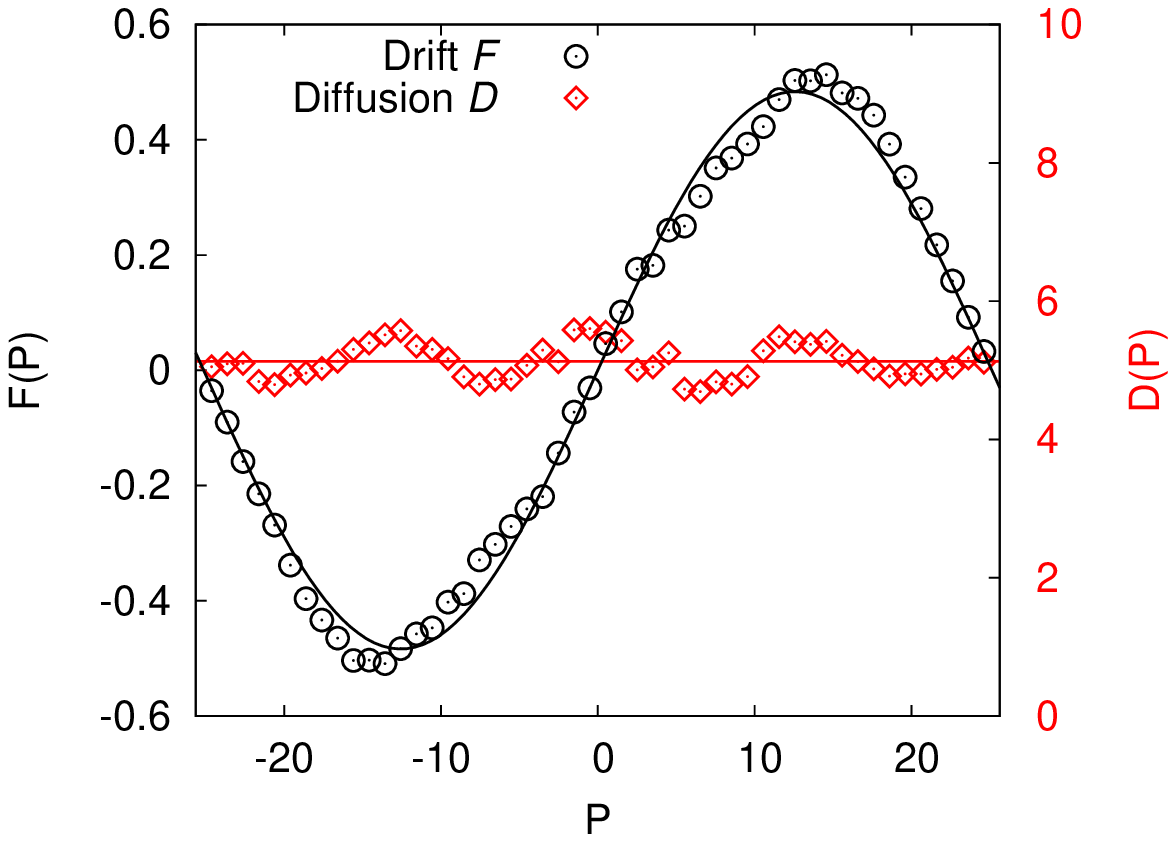}
  \caption{Drift and diffusion coefficients for the Langevin Equation describing $P$ in
  model \eqref{ham_neg2}, inferred from simulations. Left: $\beta=+0.11$; right:
  $\beta=-0.10$. In both cases $M=8$, $k=0.5$, $N=600$, $n=15$;
 integration step $\delta t=5\cdot 10^{-2}$. 
 As in Fig.~\ref{fig:fit_osc}, drifts are fitted with functions in the form \eqref{over},
 diffusivities with constant values.}
 \label{fig:fit_cpv}
 \end{figure}
The results of our extrapolation procedure are shown in Figs.~\ref{fig:fit_osc},~\ref{fig:fit_cpv}.
In all the considered cases, $D(P)$ is quite constant with
respect to the momentum, therefore equation \eqref{over} applies; indeed, $F(P)$
seems to match quite well our phenomenological prediction.
\\
Of course there is a pragmatic way to decide of the goodness of the 
above approach: compare the results from the LE and those obtained
with the exact results of the deterministic system. In Fig.~\ref{fig:autoc}
we superimpose velocity autocorrelation functions and stationary p.d.f.
obtained by stochastic simulations of the Langevin Equations to their
deterministic analogues. The similarity between the two cases is quite evident.

 \begin{figure}
 \centering
 \includegraphics[width=0.49\linewidth]{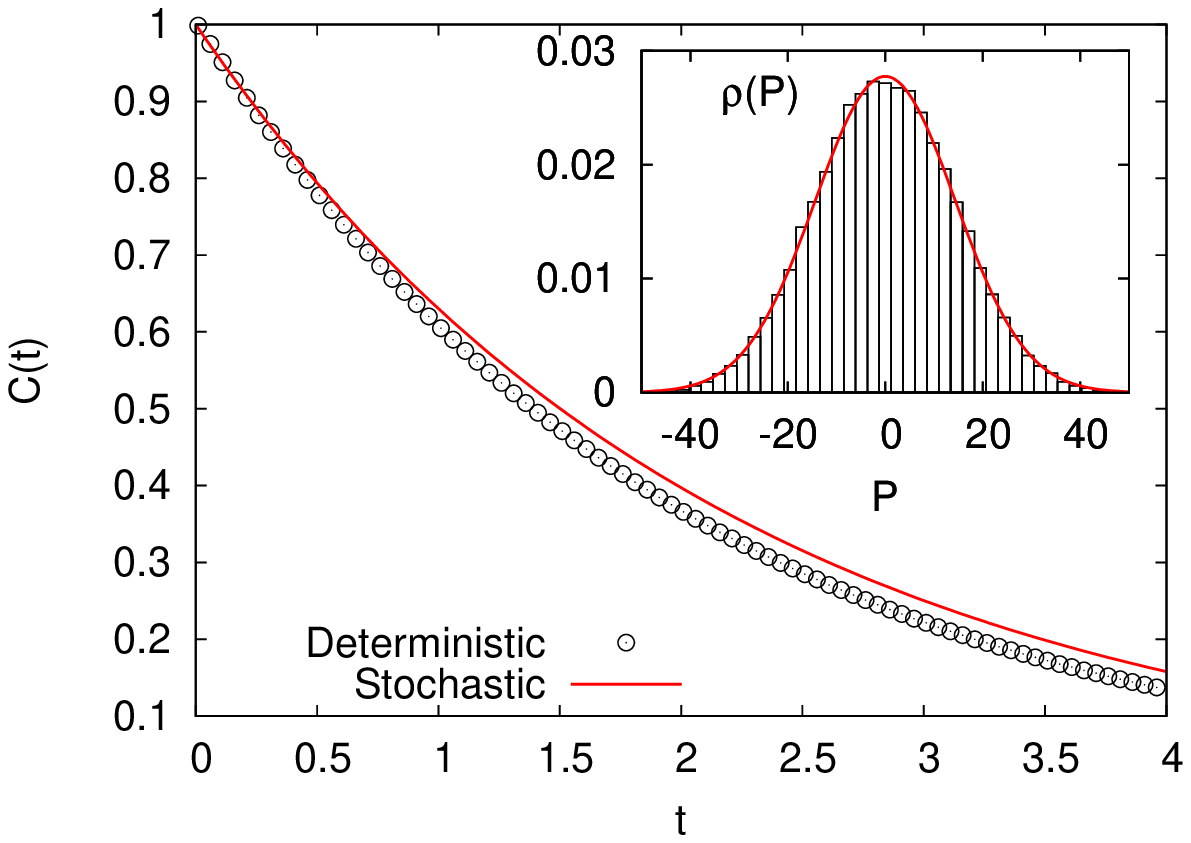}
 \includegraphics[width=0.49\linewidth]{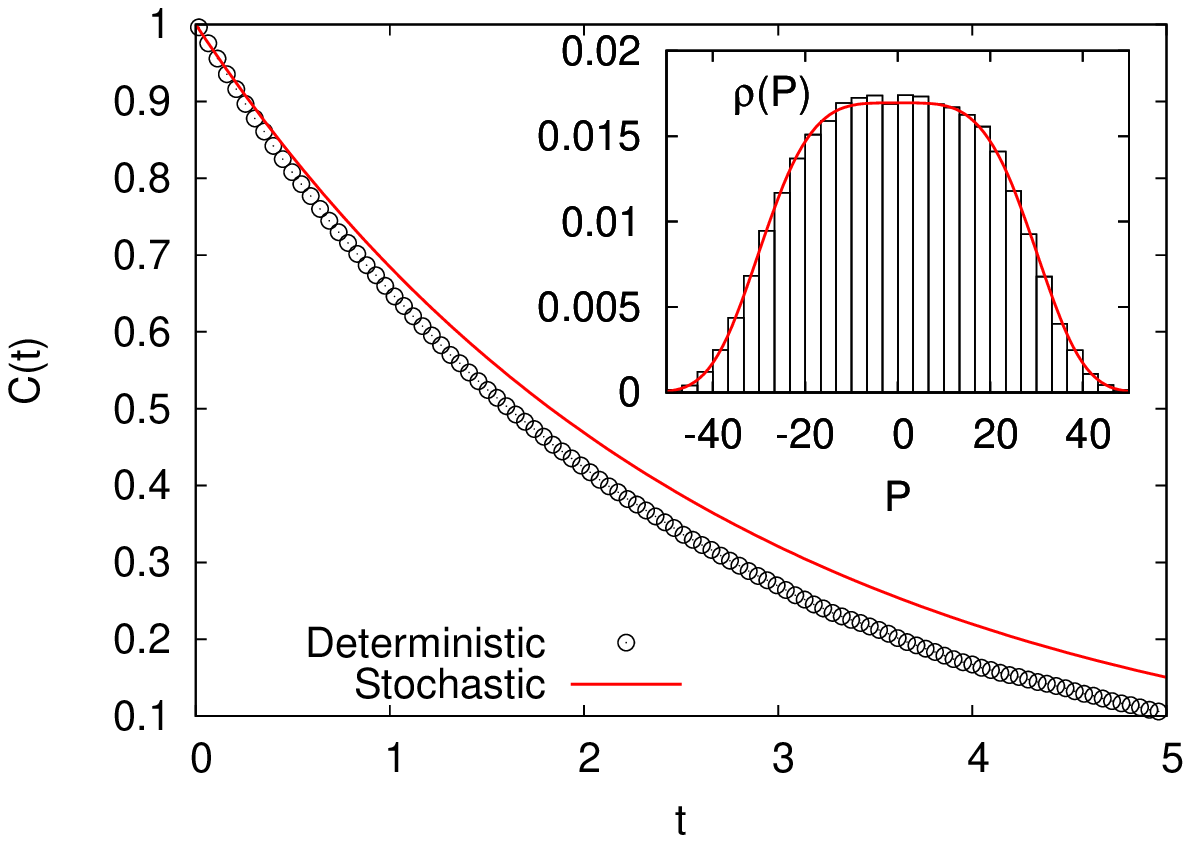}
 \includegraphics[width=0.49\linewidth]{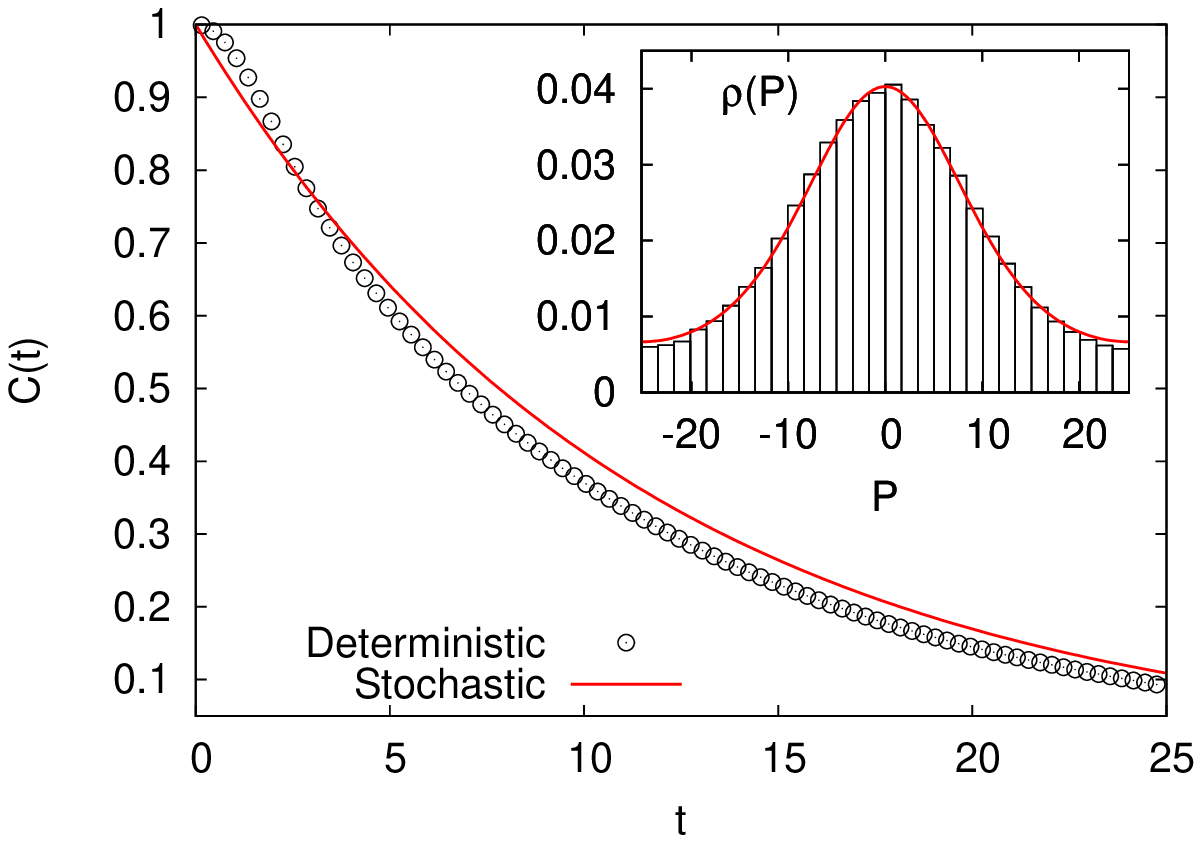}
 \includegraphics[width=0.49\linewidth]{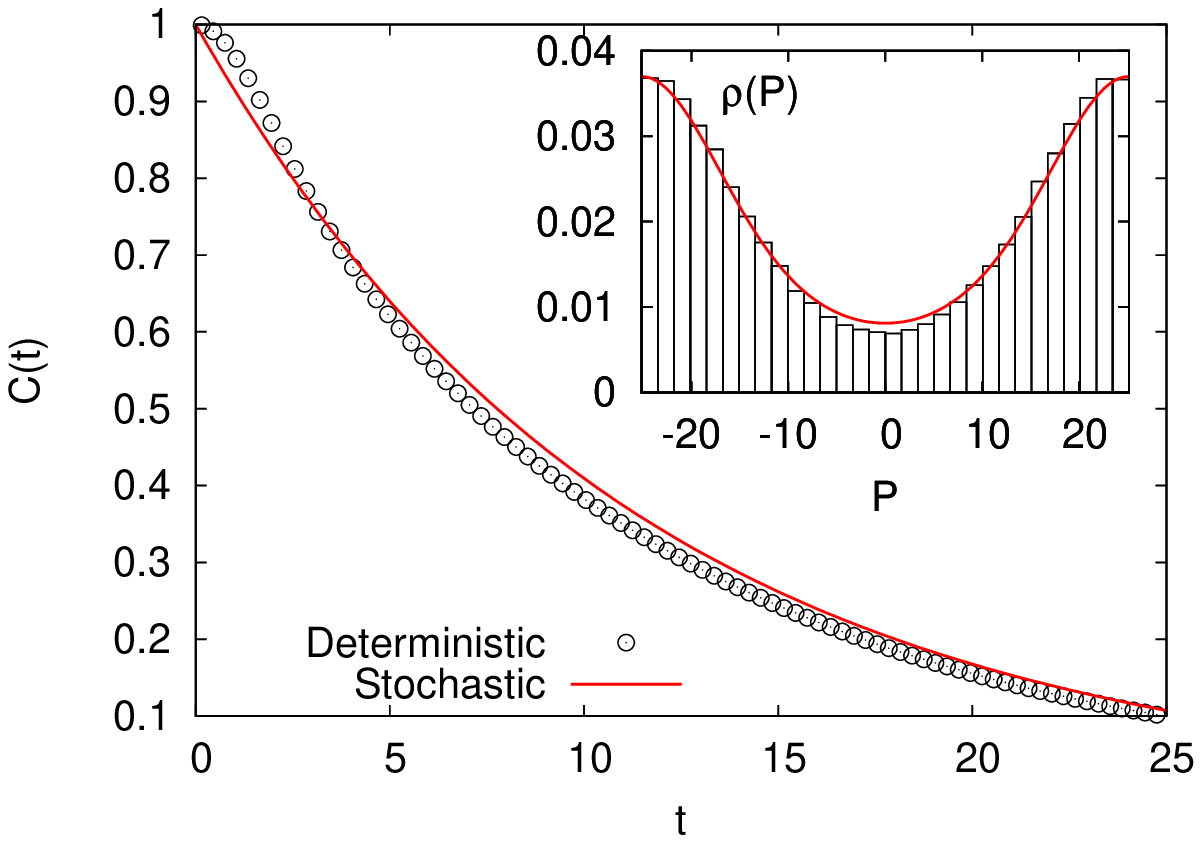}
 \caption{Autocorrelation functions for the velocity $\dot{Q}$ of the intruder in the same cases of Figs.~\ref{fig:fit_osc},~\ref{fig:fit_cpv}.
 Top-left: model \eqref{ham_osc}; top-right: model \eqref{ham_quart};
 bottom-left: model \eqref{ham_neg2} with $\beta >0$; bottom-right: model \eqref{ham_neg2} with $\beta <0$.
 Black circles represent the outcomes of
 molecular dynamics simulations, solid red lines are computed by simulating
 Langevin Equations with the previously inferred coefficients. Insets: momentum p.d.f.
 from the same deterministic (histograms) and stochastic (solid red lines) simulations.}
 \label{fig:autoc}
 \end{figure}

\subsection{Some caveats}
\label{caveats}
 
The hypothesis of a delta-correlated noise term in Eq.~\eqref{langevin}
 is quite natural (in addition it is based on an old tradition);
on the other hand one can  wonder about the possibility
of a non white noise: for instance, $\eta$ may be replaced by a stochastic process
$s$ with a non zero correlation time $\tau$, described by
$$
{ds \over dt}= -{1 \over \tau}s + c \eta \,.
$$
At first, one can say that the agreement between the statistical features
obtained with the Langevin equation and the numerical simulation
is an indirect check of the validity of the assumption on the white noise.
For a direct check one can compute the correlation function of the variable
\begin{equation} \label{zeta}
Z(t)={dP(t) \over dt} -F(P(t))\,,
\end{equation}
where $F(P)$ is determined by our fitting procedure
and $P(t)$ is obtained by the numerical simulation.
\\
In Fig.~\ref{fig:zeta} (see Appendix) we show the correlation $C(t)=\langle Z(t)Z(0) \rangle$
in a particular case: since the dynamics is deterministic, $C(t)$ must be non zero for small $t$;
on the other hand $C(t) \simeq 0$ for $t>t_*$ where $t_*$ is $O(\Delta t_*)$,
being $\Delta t_*$ the minimum value used in the fitting procedure
to determine $F(P)$ and $D(P)$.
\\
Beyond the numerical details, let us note that if
 a colored noise is present one can always describe the system
 with a Langevin equation with white noises including additional variables~\cite{villamaina2009fluctuation}.
In other words the (possible) presence of colored noises
 is nothing but one of the difficulties whose relevance had been clearly
 stressed by Onsager and Machlup~\cite{onsager}.

Let us also stress that a clear separation of the autocorrelation
time-scales of our elected degree of freedom with respect to those of
the other ones is not sufficient to apply blindly the
above procedure. This can be related again to the caveat by Onsager
and Machlup about the possible non Markovian character of the variable
used to describe the slow motion~\cite{onsager}.

\begin{figure}
 \includegraphics[width=0.49\linewidth]{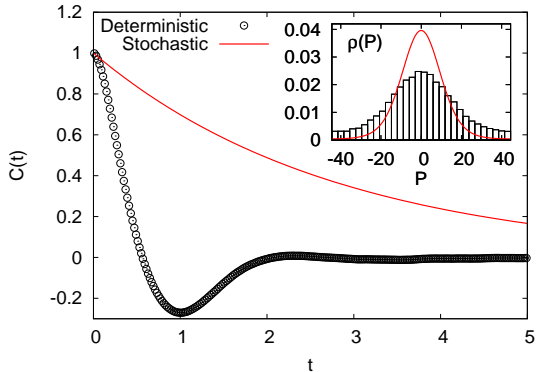}
 \caption{\label{fig:contro} Results for the Hamiltonian in
   Eq.~\eqref{ham_contro} for $M=200$, $k=2500$ $2N=2000$ and $\beta=1.04$
   and the reconstructed Langevin equation. We show the
   autocorrelations of $\dot{Q}$ in the main graph, and its probability distributions in the inset.}
\end{figure}

As far as we know there are no protocols which allow for a sure
decision, \textit{a priori}, about the Markovian character of the dynamics;
sometimes, however, some mathematical results can suggest
that our guess is wrong.  Let us consider the following system:
 \begin{equation} \label{ham_contro}
 H=1-\cos(P/\sqrt{M}) + \sum_{i=\pm 1,...,\pm N}
 [1-\cos(p_i/\sqrt{m})] + k \sum_{i=-N}^{N+1} [1-\cos(q_i-q_{i-1})],
 \quad \quad q_{-N-1}\equiv q_{N+1}\equiv 0,
 \end{equation}
where $P,q_0$ are momentum and position of the intruder. Note that
this definition violates the mass additivity discussed in the
paragraph before Eq.~\eqref{ham_quart}. In numerical simulations, for
a given choice of parameters (see Fig.~\ref{fig:contro}), we have
observed that the ratio between the typical time of the intruder and
that of the fast variable is $O(10)$, so one may hope that $P$ is
described by a $1D$ Markov model. We have used the protocol discussed
in Section III to build a Langevin equation for this model. The
results can be appreciated in Fig.~\ref{fig:contro}: we see that both
the p.d.f. of $P$ and its autocorrelation in the reconstructed model are
different from those of the real dynamics. In particular the
autocorrelation is in clear disagreement. Such a negative conclusion
could have been expected from the simple observation of the
autocorrelation of $P$ in the real dynamics: there is a range of times
$t$ where $C(t)<0$, a fact which is forbidden in $1D$ at
equilibrium~\cite{gardiner}. From the feature of numerically computed
$C(t)$ and a rigorous mathematical result one has a sure indication
that $P$ cannot be described by a $1D$ Markovian model.  It is quite
natural to guess that there exists a proper set of variables described
by a Markovian rule, but unfortunately there is not a procedure to
select such a set.

\section{Concluding remarks} 

In the present paper we introduced a practical procedure to build a
Langevin equation for slow variables from a long time series, applying
it to the data from simulations of Hamiltonian systems.  Let us note
that the use of such a protocol is not restricted to data from
numerical simulations and can be exploited also with experimental
results.  \\ In addition we show that systems with negative
temperatures does not exhibit any pathology: the dynamics of slow
variables is described by a Langevin equation whose parameters can be
computed with a well defined protocol.  Such an effective equation is
able to describe in a proper way the statistical features of the slow
variables, including their dynamics.

As a final remark, we briefly comment about the very general problem
of building models from data~\cite{chibbaro,pikovsky2017, pathak2017, ye2015}. Such an issue has attracted an
increasing interest in the recent years, in particular in the context
of the so-called Big Data paradigm and in the use of machine
learning. Surely the simplest (and well posed) problem is that of
building a model knowing the proper variables and the functional shape
of the evolution equation, see for instance~\cite{pikovsky2017}. The 
task is more difficult when one knows the proper variables while the shape of the model is
unknown~\cite{pathak2017}. The most ambitious problem is that of building a model
using just data, without any \textit{a priori} assumption about the structure
of the equations and the relevant variables. An approach inspired to
Takens has been used to write down evolution equations in ecological systems \cite{ye2015};
however such a procedure only works
in low-dimensional systems~\cite{cecconi2012}. The
building of a LE, discussed here, is part of these challenges to
extract models from data: a pragmatic attempt based on physical
intuition and numerical treatment.

\begin{ack}
We acknowledge useful discussions and correspondence with A. Cavagna, A. Pikovsky and A. Sarracino.
\end{ack}

\appendix

\section{Details on the  numerical protocol}

\begin{figure}
 \centering
 \includegraphics[width=0.49\linewidth]{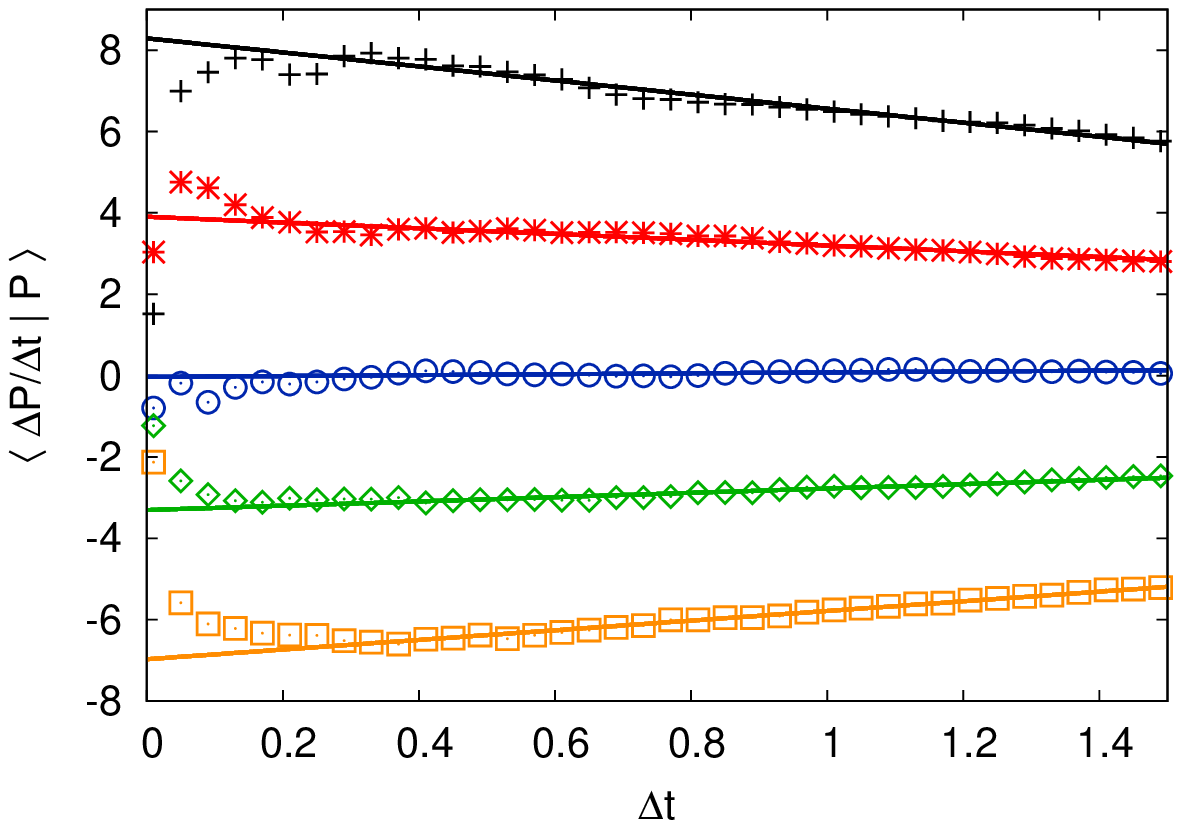}
 \includegraphics[width=0.49\linewidth]{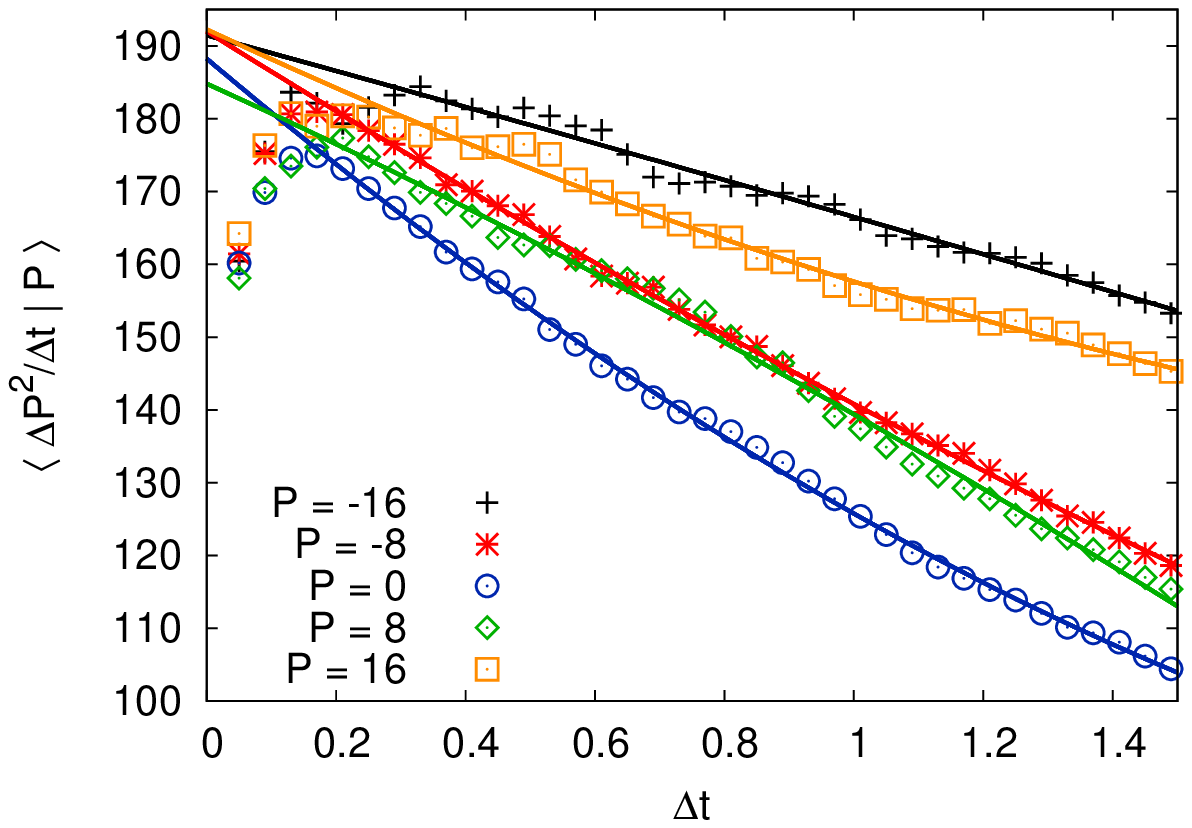}
 \caption{Extrapolating limits \eqref{coeff} for some values of $P$ (in the
 same conditions of Fig. \ref{fig:fit_osc}). Drift (left) and diffusion (right) coefficients are
 inferred by considering linear and  parabolic fits, respectively.
 The considered data interval for each fit is, in this case, $\Delta t \in [0.25, 1.5]$. }
 \label{fig:extr}
 \end{figure}

In this Appendix we discuss the numerical method we employ to
infer drift and diffusion coefficients for the Langevin Equations of the heavy
particle's momentum. The idea is to perform molecular dynamics simulations of
the Hamiltonian systems discussed in Section \ref{sec:ham}, in order to compute
the limits \eqref{coeff} from averages on long time series of data.
\\
All simulations are prepared in equilibrium initial conditions. This is particularly relevant
for the harmonic chain case, since this Hamiltonian system is integrable and by no chance 
it can reach the proper equilibrium p.d.f. starting from out-of-equilibrium conditions;
in this case one has to properly distribute the total energy among the normal modes
of the chain.
\\
Let $I=(-\tilde{P},\tilde{P})$ be a typical range for the momentum of the heavy
particle, $P(t)$, and let us divide it into $n$ intervals 
$I_1, I_2, ..., I_n$ of equal lengths. During the simulation, our algorithm periodically checks
for what $j$ (if any) the relation $P(t)\in I_j$ holds; then it stores the values
of $P(t+\Delta t)- P(t)$ for several $\Delta t$. 
In order to avoid correlations, the delay between two measures has been chosen to be at least $2 \tau$,
where $\tau$ is the velocity autocorrelation time of the intruder. At the end of the process, conditional averages
on the r.h.s. of Eqs.~\eqref{coeff} are computed as functions of both $j$ and $\Delta t$, and the
limit $\Delta t \to 0$ can be inferred.
\\
Let us note that the mathematical limit  $\Delta t \to 0$ in the definitions of drift
and diffusion coefficients must be interpreted in a proper physical way.
It is quite simple to show that the description of a deterministic system
in terms of a Langevin equation cannot be completely accurate at any time scale,
but only  at times larger than a certain characteristic threshold.
\\
This can be easily understood noting that in a deterministic system, 
for any normalized correlation function,
$$
C_D(t)=1- {t^2 \over \tau_D^2}+O(t^3)\,;
$$
on the contrary, for a system whose evolution is ruled by a LE one has
$$
C_L(t)=1- {t \over \tau_c}+O(t^2) \,.
$$
From the comparison of the two correlation functions it follows that 
the Markovian approximation can be valid only for 
$$
t > t_*=O\Bigl({\tau_D^2 \over \tau_c}\Bigr)\,.
$$
As an example we can mention the case of an intruder in the harmonic chain seen in
section \ref{sec:ham}, for which some exact results can be used \cite{zwanzig}:
here $\tau_D=\sqrt{M/k}$ and $\tau_c=M/\sqrt{4k m}$, so that the previous
approximation only applies for $t > 2\sqrt{m/k}$.
\\
Of course in general it is not simple  to find \textit{a priori} $t_*$ and therefore determine
the minimum acceptable value of $\Delta t$.
In our numerical computations of the drift we adopt a rather pragmatic
approach:  for a given $P$ we use different $\Delta t$ and then extrapolate the result using
values of $\Delta t$ which are not too small. Fig. \ref{fig:extr} shows a typical situation:
fortunately one has an easy natural way to perform the extrapolation, i.e. using
a polynomial function to fit the data and considering its value for
$\Delta t=0$. In our analysis we employ linear fits for the drift and
parabolic ones for the diffusion.

\begin{figure}
 \centering
 \includegraphics[width=0.49\linewidth]{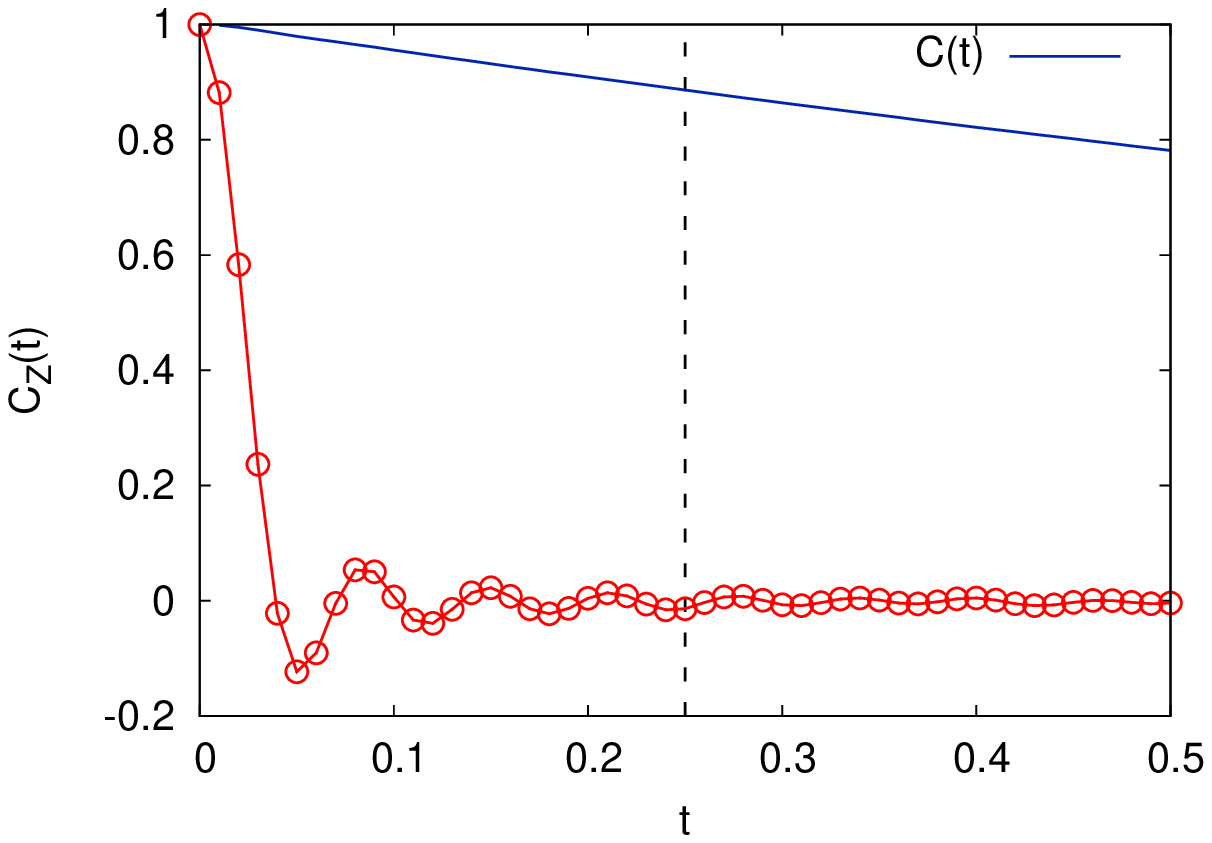}
 \caption{Autocorrelation of the $Z(t)$ function defined by Eq.~\eqref{zeta}
 for model \eqref{ham_osc}  (red circles), in the same conditions of
 Fig.~\ref{fig:extr}. Velocity correlation function $C(t)$ (blue
 solid line) is shown for comparison; the minimum time interval that has been used 
 in extrapolating the limits \eqref{coeff} is, in this case, $\Delta t_*=0.25$ (dashed vertical line).}
 \label{fig:zeta}
\end{figure}
With the aim of verifying that the noise is fairly approximated by a
delta-correlated process, we compute the autocorrelation of the
quantity $Z(t)$ defined in~\eqref{zeta}, whose plot is shown in
Fig.~\ref{fig:zeta}. We see that indeed the ``noise'' $Z(t)$ loses
memory in a time clearly smaller than $\Delta t_*$, the minimum value of $\Delta t$ we consider for the
extrapolations (see dashed line), and certainly much smaller than the
correlation time of the slow variable $P$.

To check the goodness of our extrapolation method one could, of course,
compare the results to the analytical predictions valid for $M/m\rightarrow \infty$
in the thermodynamical limit. In this case, anyway, the resulting deviation of the
measured values from the theoretical ones would be affected not only by the actual 
errors in the extrapolation procedure, but also by the fact that considering $P(t)$ as
a stochastic process is by itself an approximation. At least in the case of
the harmonic chain, however, we can do better than this: since the conditional
p.d.f of $P$ is known, there is an easy way to compute the Langevin parameters
from data without performing the limit $\Delta t \rightarrow 0$, and we can compare
these values to the results of the previous method in order to estimate the precision
of the extrapolation.
\\
\begin{figure}
 \centering
 \includegraphics[width=0.49\linewidth]{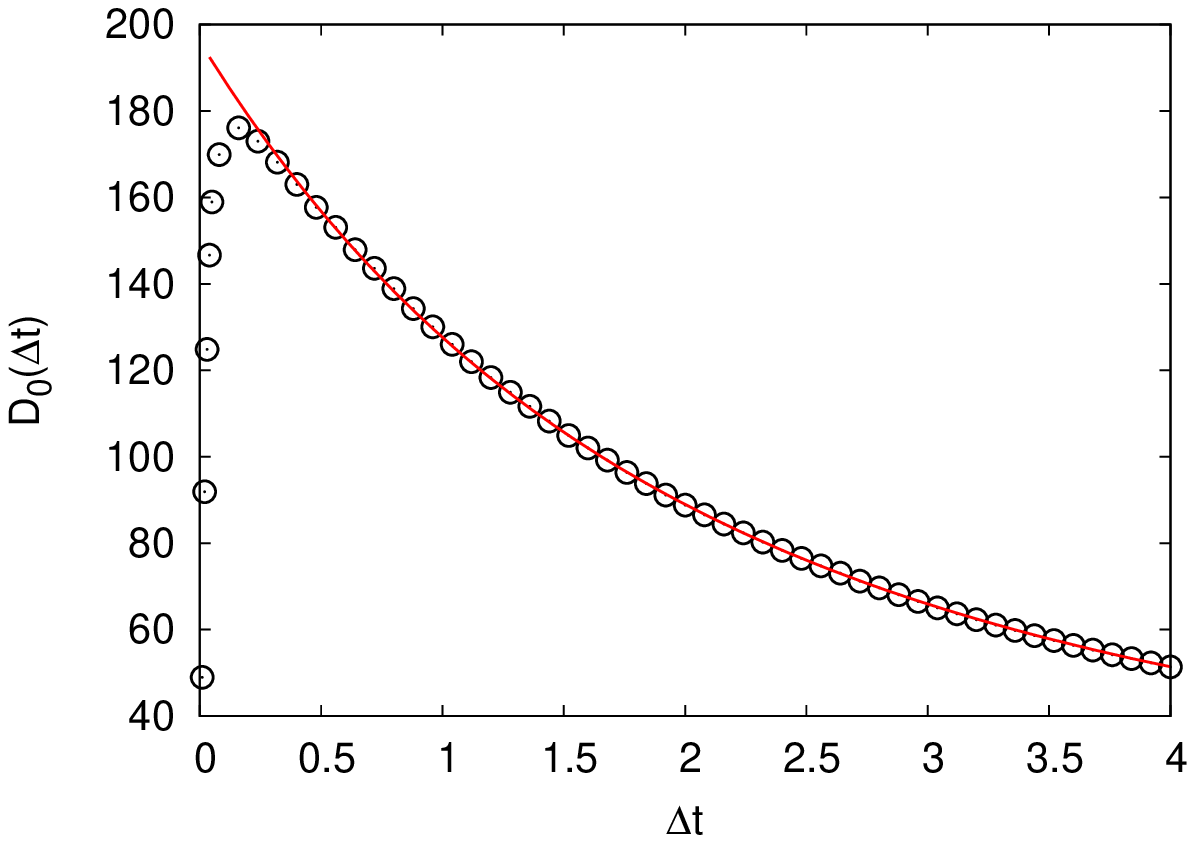}
 \caption{Evolution of $D_0(\Delta t)$ with the time interval. The red line is the 
 result of a fit with the functional form \eqref{d0}.}
 \label{fig:minimum}
 \end{figure}
To this end, let us note that since the conditional p.d.f. of $P$ is given by \cite{rubin60}: 
\begin{equation}
f(P(t+\Delta t)|P(t))=\frac{1}{\sqrt{2\pi k_B T M(1-C(\Delta t)^2)}}\exp\left\{-\frac{[P(t+\Delta t)-P(t) C(\Delta t)]^2}{2k_BTM(1-C(\Delta t)^2)}\right\}\,, 
\end{equation}
we can explicitly compute the averages in Eq.~\eqref{coeff}, assuming that the
Markovian limit holds and that the autocorrelation function actually verifies
$C(\Delta t)=\exp(-\Delta t/\tau)$ for some $\tau$.
For the diffusion term we get
\begin{equation}\label{diff_dt}
 {1 \over \Delta t} \langle \Delta P(\Delta t)^2 | P(t)=P \rangle =D_0(\Delta t) + D_1(\Delta t)P^2\,
\end{equation}
where 
\begin{equation}\label{d0}
 D_0(\Delta t)=\frac{k_BTM}{\Delta t} \left(1-e^{-2\Delta t/\tau} \right)\,,\quad \quad D_1(\Delta t)=\frac{\left(1-e^{-\Delta t/\tau} \right)^2}{\Delta t}P^2\,.
\end{equation}
\\
We can fit our data with the previous formula: in particular, from the fit of 
$D_0(\Delta t)$ (Fig.~\ref{fig:minimum}), we infer both $T$ and $\tau$ and
calculate the corresponding values of drift and diffusion for Brownian motion \cite{gardiner},
$$
F(P)=-P/\tau\,,\quad \quad D(P)=2Mk_BT/\tau\,.
$$
The resulting values and the previously extrapolated coefficients differ by less than 3\%,
which is a quite satisfactory precision for our qualitative analysis.

\section*{Bibliography}

\bibliographystyle{iopart-num}
\bibliography{biblio}
\end{document}